\documentclass[conference]{IEEEtran}

\usepackage{float}
\usepackage{multirow}
\usepackage{booktabs}
\usepackage{multirow}
\usepackage{siunitx}
\IEEEoverridecommandlockouts
\usepackage{cite}
\usepackage{amsmath,amssymb,amsfonts,amsthm,bm}
\usepackage{algorithmic}
\usepackage{graphicx}
\usepackage{textcomp}
\usepackage{xcolor}
\usepackage[utf8]{inputenc}
\usepackage{flafter}
\usepackage[font=scriptsize,skip=5pt]{caption}
\usepackage{float}
\usepackage{comment}
\usepackage{mathtools}
\usepackage{subcaption}
\usepackage{pgfplots}
\usepackage{ifthen}

\usepackage{tikz}
\usetikzlibrary{calc,math}

\usetikzlibrary{calc,math}

\begin{document}
\title{Performance Evaluation of  Video Streaming Applications with Target Wake Time in Wi-Fi 6}

\author{\IEEEauthorblockN{Govind Rajendran\IEEEauthorrefmark{1}, Rishabh Roy\IEEEauthorrefmark{2}, Preyas Hathi\IEEEauthorrefmark{3}, Nadeem Akhtar\IEEEauthorrefmark{3} and Samar Agnihotri\IEEEauthorrefmark{1}}
	
\IEEEauthorblockA{
 		\IEEEauthorrefmark{1} Indian Institute of Technology Mandi,  India\\
 	\IEEEauthorrefmark{3}	  Arista Networks \\	
 	\IEEEauthorrefmark{2} Indian Institute of Technology Dharwad, India \\
 		Email: \{\IEEEauthorrefmark{1}s21010\}@students.iitmandi.ac.in,
 		 \{\IEEEauthorrefmark{1}samar\}@iitmandi.ac.in,
 		\{\IEEEauthorrefmark{3}preyas.hathi, \IEEEauthorrefmark{3}nadeem.akhtar\}@arista.com}
 		\{\IEEEauthorrefmark{2}201082002\}@iitdh.ac.in
}

\maketitle

\begin{abstract}
    The Target Wake Time (TWT) feature, introduced in Wi-Fi 6, was primarily meant as an advanced power save mechanism. However, it has some interesting applications in scheduling and resource allocation. TWT-based resource allocation can be used to improve the user experience for certain applications, e.g., VoIP, IoT, video streaming, etc. In this work, we analyze the packet arrival pattern for streaming traffic and develop a synthetic video streaming traffic generator that mimics real-world streaming traffic. We propose a two-stage approach where we calculate the TWT duty cycle in the first step. In the subsequent step, we determine the Multiplication Factor (MF), which jointly dictates the required TWT schedule for the synthetic traffic model. Initial testing shows that key QoS metrics can be met for sustained performance of synthetic traffic upon enabling TWT, even in the presence of peak background congestion in the network.
\end{abstract}
\section{Introduction}
Wireless local area network (WLAN) technology, popularly referred to as Wi-Fi, has a major impact in our day to day lives for high speed wireless access because of  its low cost and ease of deployment. In homes, corporate offices and public places such as stadiums, railway stations, airports etc, Wi-Fi provides high-speed connectivity to the Internet. However, due to the use of shared, unlicensed spectrum for Wi-Fi, congestion is a major problem, leading to network performance deterioration both in terms of throughput and latency. WLAN client devices are typically portable, equipped with power limited batteries. As a consequence, power saving mechanisms play an important role in battery life longevity and device performance. Wi-Fi 6 was introduced, based on the IEEE 802.11ax version of the WLAN standard, to improve the end user Quality of Service (QoS) by efficient use of the frequency spectrum in dense WLAN deployments and reduced power consumption for power limited devices. Wi-Fi 6 introduces 1024-QAM, orthogonal frequency division multiple access (OFDMA), 8-stream multi user-multiple input multiple output (MU-MIMO), spatial reuse to improve end user performance and increase spectral efficiency even in dense scenarios\cite{Standard,Latency1}. In addition to that, to improve existing power management methods, a new power save mechanism called Target Wake Time (TWT) is introduced in Wi-Fi 6. The TWT feature is primarily meant as an advanced power save mechanism, which may not be used in mission critical scenarios where the client needs to stay awake for the entire duration of operation. However, it has some interesting applications in scheduling and resource allocation\cite{Bellata-TWT-Scheduled-Access}. Wi-Fi 6 supports two types of TWT mechanisms: (i) Individual, and (ii) Broadcast. In the Individual TWT mode, AP creates service schedules on a per-client basis. A TWT schedule consists of wake-up and sleep intervals where clients can transmit/receive frames in the wake-up intervals and turn off their transceivers and go to a power-saving mode during the sleep intervals. It is also possible to group clients and create group-level service schedules using Broadcast TWT mode. With this, APs can create a quasi-TDMA type of resource scheduling, on top of OFDMA \cite{DBLP:Bellata-journals/corr/abs-1811-00957,DBLP:Uplink-Resource-Allocation} and MU-MIMO\cite{ONLINE:Arista-White-Paper-MU-MIMO}. With pre-defined wake-up times for each client, TWT can help reduce channel contention and improve airtime utilization since only a subset of the associated clients are expected to be active at a given time. In a managed, multi-AP Wi-Fi network, an additional level of coordination can be applied for co-channel neighbor APs to reduce contention across APs. The challenge is to choose the appropriate TWT mode and the schedules to address the needs of clients and applications. One important point to remember is that the opportunistic channel access method, CSMA/CA rule still applies during the wake-up intervals and the AP has to contend for the TWT opportunities in an environment where clients that do not understand TWT or are not assigned TWT schedules are also present.

TWT-based resource allocation can be used to improve the user experience for certain applications, e.g. VoIP. Note that VoIP follows the ON-OFF model, corresponding to active talking and silent periods \cite{ONLINE:WLAN-TGax-Evaluation-methodology}. Also note that the VoIP payload size during the active and silent periods is fixed (depending on the codec used). For such an application, a TWT schedule can be created considering the traffic model features. If there is a dedicated Wi-Fi network for VoIP, then clients connecting to this Wi-Fi network can be grouped together and assigned appropriate TWT schedules. Same approach can be applied to IoT applications, for example sensors which generate telemetry at periodic intervals. TWT can also be used for real-time applications like video streaming. In this case, TWT is used to provide guaranteed airtime to users. The process requires characterizing the traffic model first and then generating the schedule to fit the packet arrival process. While there are well-documented statistical models for video streaming\cite{ONLINE:WLAN-TGax-Evaluation-methodology}, most of the popular applications like YouTube  \cite{mondal2017candid}, Netflix, etc. have proprietary mechanisms in place to adapt the video stream to the link quality, which necessitates characterizing the traffic flows and optimizing the TWT schedules separately for each of these applications.

In this work we present the performance of streaming applications running on \textit{a single} TWT enabled client operating in a congested network and the impact of enabling TWT on the network's  overall performance.
The main contributions of this paper can be summarized as follows:
\begin{itemize}
    \item We create a synthetic video streaming traffic generator model that approximates real-world video streaming and also define some relevant QoS metrics for performance evaluation, like ‘average throughput’ and ‘buffer health status’.
    \item We propose a two-stage approach to find a TWT schedule satisfying the QoS requirements for streaming traffic, thereby improving end-user experience.
\end{itemize}

\section{Application Traffic Model and QoS Metrics}
We pick video streaming application because the bulk of today’s network traffic consists of streaming and video-on-demand (VOD) services\cite{streaming_data}. We analyze the packet arrival pattern for video streaming traffic and develop a synthetic streaming traffic generator that in essence mimics real-world streaming traffic. To analyze the performance of TWT on streaming services, we need to define appropriate Quality of Service (QoS) metrics that will quantitatively measure the system performance.
\label{for result}
\subsection{Synthetic traffic model}
Most streaming services use DASH (Dynamic Adaptive Streaming over HTTP)\cite{dash-cloudflair} as an application layer protocol to stream videos. In DASH, the content server stores a video by breaking it down into \textit{bursts}, usually worth a few seconds of video time. Initially, when the video starts playing, the server sends these bursts at a high rate in order to fill up the playback buffer with $45$-$60$ seconds worth of video time (known as \textit{Startup phase}). The server then tops up the buffer (known as \textit{Bursty phase}) periodically. In the bursty phase, bursts generally carry $2$-$10$ seconds worth of video data \cite{zhang2017modeling}. This buffered approach allows the streaming service to handle intermittent network issues without compromising video quality. By not buffering the whole video at once, DASH also minimizes the data loss in case the user decides to switch to a different video. DASH also stores the video at different picture resolutions (different size bursts) and dynamically switches among them based on the client's bandwidth. In DASH, it is the client who measures the available bandwidth as:  
\begin{center}
    $\text{Available bandwidth}=\frac{\text{Size of the last burst}}{\text{Time taken to serve the last burst}}$
\end{center}
If the network cannot accommodate the current bitrate, the client requests for lower bit rate video from the server.\cite{7057917}



 \begin{table}[!ht] 
 
\centering
\begin{tabular}{ |p{1.2 cm}|p{4 cm}|}
 \hline
  \textbf{Parameter} &\textbf{Description} \\ [0.5ex] 
 \hline
  Frame Size & Weibull distribution with parameters, $k=0.8099$ and
  \newline
  $\lambda = (\frac{6950\times \text{\rm Video bit rate}}{2})$ \cite{WLAN-TGax-Evaluation-methodology}  \\ 
  \hline
  Inter-burst time & Truncated normal distribution with mean of $6$ seconds and variance of $1.8$ seconds and lies within $2$-$10$s \cite{dash-cloudflair} \\
  \hline
  Frame Rate & $30$ frames per second\\
 \hline
 
\end{tabular}
\caption{VBR Traffic Model}
\label{TABLE:Synthetic-Traffic-Parameters}
\end{table}

We consider two synthetic traffic models, Constant Bit Rate (CBR) and the Variable Bit Rate (VBR). Both models use TCP as the transport layer protocol. Traffic for both models is sent over a multi-hop global network to incorporate real network delays. We define the following parameters to characterize the streaming traffic models in their bursty phase:
\begin{itemize}
    \item \textbf{Inter-burst time: } It is the time duration, in seconds, between two consecutive bursts.
    \item \textbf{Burst Size: } The size of each individual burst, expressed in Megabytes (MB).
\end{itemize}

\subsubsection{CBR Model}
In the CBR model, a burst of fixed size, corresponding to the video bit rate, is generated every set amount of inter-burst time. To approximate the bursty nature of DASH protocol, the inter-burst time is set at $6$ seconds, which is the mean value of the inter-burst time distribution with parameters mentioned in the Table \ref{TABLE:Synthetic-Traffic-Parameters}. It acts as our baseline traffic model whose QoS requirements need to be satisfied, before we test with the more dynamic VBR model.
\label{CBR}
\subsubsection{VBR Model}
In order to closely mimic the bursty state of real streaming, we consider the VBR model where its traffic parameters are mentioned in Table \ref{TABLE:Synthetic-Traffic-Parameters}. Inter-burst time is sampled and multiplied with the required frame rate (i.e., the number of frames per second) to get the expected number of frames for that duration. We then sample the frame size for each frame to be transmitted and get the total size of the burst. 

\label{VBR}
\subsubsection{The need for a synthetic streaming model}
Though the traffic of real streaming follows the bursty trend, the exact pattern with reference to inter-burst time and burst size is dependent on multiple factors including picture quality, genre of video playing, rate and magnitude of change in content across frames, compression algorithm used by the streaming service, transport layer protocol used (TCP, UDP or QUIC) etc.
The synthetic streaming model described incorporates video frame generation from \cite{WLAN-TGax-Evaluation-methodology} and inter-burst time generated from the statistics provided in \cite{dash-cloudflair}, \cite{zhang2017modeling}. We use this synthetic model rather than emulating traffic patterns of a real video as the patterns of each video are different and by using the synthetic streaming model we are able to incorporate this variability while also keeping the trend of bursty traffic. It also gives us a greater control over the experimental parameters like bit-rate of video and duration and allows us to quantitatively compare performance while also capturing sender side statistics.
\label{Traffic description}
\subsection{QoS parameters}
The perceived experience of the end user while using an application is the true measure of the Quality of Service provided by the network. Metrics such as the Mean Opinion Score are subjective and often take a lot of human time and effort to rate the application’s user experience. Instead, objective measurements of certain quantifiable factors which are known to contribute to the experience of an application is easier, involves less human effort, and allows us to quickly measure the performance of the network. The following QoS metrics are relevant  for evaluating the performance of streaming:
\begin{itemize}
    \item \textbf{Throughput:} Application layer throughput is one of the most basic yet important metrics that determine performance. If a streaming service's throughput requirement is not met it will lead to buffering or quality change in the video.
    \item \textbf{Buffer Health Status:} As described above, DASH uses a buffer to maintain a threshold of video playback time. The buffer at the client stores the incoming burst to facilitate smooth playback of the video. When the traffic falls below this threshold i.e. a buffer underrun occurs where the rate of consumption is higher than the rate at which traffic enters the buffer, DASH adapts and changes the video to a lower resolution to maintain the playback time. Change in video rate severely affects the experience of watching the video, hence maintaining the buffer health is one of the most important QoS metrics for streaming
    \cite{ghobadi2012trickle}.
    \item \textbf{Throughput Variation:}  While we may meet the average throughput requirements, there might be large fluctuations in the instantaneous throughput. Variation in throughput is not an absolute metric of performance, but it can severely affect the buffer health due to the bursty nature of streaming traffic and potentially be one of the contributing factors of QoS failure.
\end{itemize}
\label{QoS}
\label{Application Traffic Model and QoS Metrics}
\section{TWT Background} In this work, we consider the Individual TWT operation mode where the Wi-Fi 6 AP negotiates TWT agreements with Wi-Fi 6 capable devices at a per-client basis. Individual Target Wake Time comprises of the following parameters (refer to Fig. \ref{fig:Individual-TWT}): 
\begin{figure}[!t]
\includegraphics[width=0.5\textwidth]{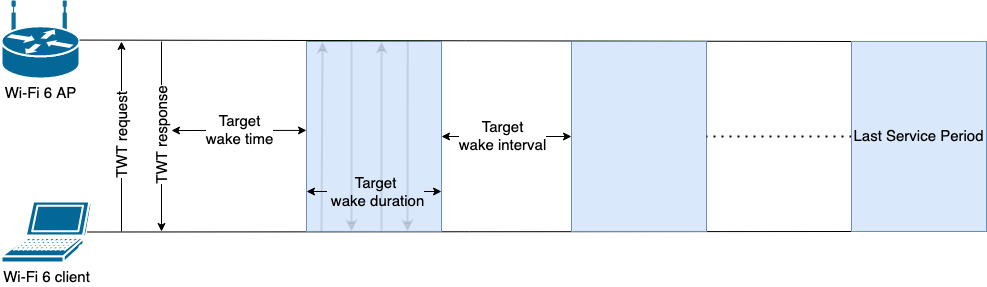}
\caption{An illustration of Individual TWT.}
\label{fig:Individual-TWT}
\centering
\end{figure}

\begin{itemize}
\item \textbf{Target Wake Duration: } It is the minimum amount of time for which the TWT-enabled client is awake and transmits/receives data. We denote Target Wake Duration as TWT-SP. Each TWT-SP consists of multiple Wi-Fi 6 frame transmissions.
\item \textbf{Target Wake Interval: }  It is the time difference between two consecutive TWT-SPs where the client can go to sleep. We denote Target Wake Interval as TWT-WI. TWT-WI can be periodic (known as ‘implicit’ TWT) or aperiodic (known as ‘explicit’ TWT) depending on the type of agreement AP negotiates with the client. 
\item \textbf{Target Wake Time: } It is the time offset from the TWT negotiation till the starting of the first TWT-SP.

In this work, we use the ‘Unannounced’ setting inside the Individual TWT agreement, where the AP expects the clients to be awake throughout a service period and the clients need not send any PS-POLL or APSD trigger frames to AP. We also consider the negotiated TWT to be an ‘implicit’ type of agreement where a TWT-enabled client wakes up and goes back to its sleep state periodically.
\end{itemize}

\subsection{The TWT Duty Cycle and Multiplication Factor}
We define TWT \emph{Duty Cycle} to be the percentage of time a TWT-enabled client is awake throughout the TWT schedule. It is calculated as:
\(\frac{(TWT-SP)}{(TWT-SP+TWT-WI)}\times100\%\) 
\newline
It essentially dictates guaranteed wake time for the TWT-enabled client to transmit/receive frames. How often the TWT-enabled client wakes up in a particular TWT Duty Cycle is governed by another parameter, referred to as \emph{Multiplication Factor} (MF) i.e. a TWT-enabled client can wake up more (respectively, less) frequently if it uses a higher (respectively, lower) MF, while maintaining the same TWT Duty cycle.

\subsection{Calculation of TWT Schedules}
The IEEE 802.11ax standard defines the method to calculate the TWT schedules which consist of the parameters mentioned in \cite{Standard}. The implementation of TWT puts an upper limit on the duration of TWT-SP, precisely $65535$ $\mu$S. As a direct consequence, it impacts the TWT schedules, Duty Cycles and MF calculations. The TWT schedule calculated with TWT-SP = $65535$ $\mu$S corresponds to $\text{MF} = 1$ for a particular TWT duty cycle. We derive TWT schedules corresponding to a higher MF value by dividing both the TWT-SP and the TWT-WI obtained from MF = 1, with the desired MF (approximately).

\section{Experimental Setup and results}
\begin{figure}[t!]
\includegraphics[width=0.5\textwidth]{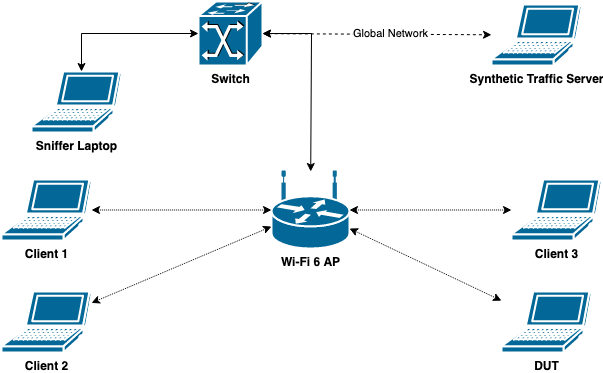}
\caption{Experimental setup}
\label{fig:testbed}
\centering
\end{figure}
The test bed used to perform our experiments is shown in the Fig. \ref{fig:testbed}. We use an enterprise-grade Wi-Fi 6 capable AP which supports individual TWT. The AP is configured to use a 20MHz channel. AP is the central node that receives and sends all data, provides a centralized way to solicit the TWT commands and keep track of the performance of the network. The AP also doubles up as an iPerf server to generate downlink TCP multistream traffic to load the network. We use multiple  non-TWT Wi-Fi 6 clients namely Client 1, Client 2, Client 3 that act as iPerf clients; to mimic real deployments, the clients have different RSSI values. The device under test (DUT) is a TWT enabled client to which we send synthetic traffic from the synthetic traffic server through a multi-hop network and evaluate QoS. The Table \ref{table:Set-up-parameters} briefly describes the physical parameters that are relevant to our experimentation.

\begin{table}[!ht]
\centering
\begin{tabular}{ |p{1 cm}|p{0.7 cm}|p{2 cm}|p{2.2 cm}|} 
 \hline
 Client Index &
RSSI (dBm) &
Peak Standalone Throughput&
Multi-client Scenario Throughput \\
\hline
  1 &
-46 &
63.5 Mbps &
14.3 Mbps
 \\ 
 \hline
  2 &
-45 &
75.4 Mbps &
23 Mbps\\
\hline
 3 &
-37 &
163 Mbps &
17.6 Mbps
 \\
 
 \hline
 4 (DUT)&
-36&
95 Mbps &
21.3 Mbps 
\\
 \hline
\end{tabular}
\caption{Setup Parameters.}
\label{table:Set-up-parameters}
\end{table}

\subsection{Methodology}
In this subsection, we discuss the methodology to obtain the appropriate TWT schedule for the synthetic traffic model which guarantees the required QoS requirements mentioned in Section~\ref{QoS}. We follow a two-stage approach where at each stage, we satisfy one of the two important QoS requirements for this work, i.e., average throughput and buffer health status, for a specified video bit-rate. The procedure works as follows:

    \subsubsection{Phase 1 } In this phase, we obtain the smallest TWT duty cycle required for a specified bit-rate video stream. The metric used for this phase is average throughput; the goal is to find the TWT duty cycle that satisfies the QoS requirement. Towards this, we increase the duty cycle in steps of $5$\% and observe the corresponding average throughput values for a single TCP stream iPerf session from the AP to the DUT. The rationale behind using iPerf here is that it is one of the most common, open-source, tools for measuring network performance. This gives us an upper bound on the average throughput that can be achieved by the system for the given TWT duty cycle in an unloaded network.
    \begin{figure}[!ht]
\includegraphics[width=0.5\textwidth]{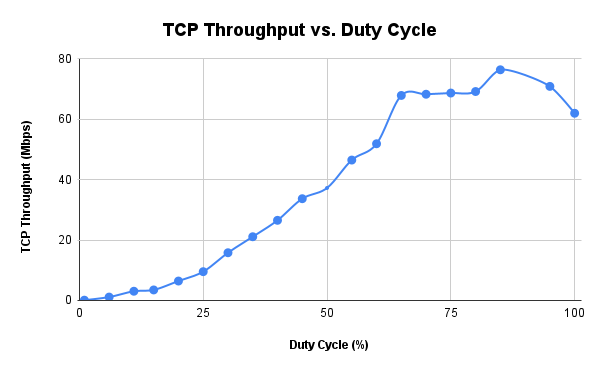}
\caption{Single TCP stream throughput variation with different TWT duty cycles}
\label{fig:Throughput Graph}
\centering
\end{figure}
 From Fig. \ref{fig:Throughput Graph} we can look-up the required smallest TWT duty cycle value to achieve the average throughput requirement for the DUT.
    \label{phase1}
    \subsubsection{Phase 2 } In this phase, the QoS metric to meet is buffer health status. {In our experimentation the buffer health status though being a client side metric, is actually measured at the server for the following reasons:
    \begin{itemize}
        \item The client is oblivious to the inter-burst time as it is sampled from Table \ref{TABLE:Synthetic-Traffic-Parameters} by the server.
        \item Since TCP is used (end to end transmission is guaranteed) the server knows exactly when the burst has been served out. Coupled with the fact that the server already knows the inter-burst time it makes it convenient to measure the buffer health status from the server side.
    \end{itemize}

    This metric places a stricter demand on throughput; in order to maintain the buffer health, i.e., avoid buffer underruns, we need the average throughput during the inter-burst time to be larger than the video bit rate. Towards this, we start with the smallest TWT duty cycle obtained from Phase $1$ and observe the instantaneous throughput variation for different MF. From Fig. \ref{fig:Instantaneous Graph}, we observe that when MF is low, there is a large number of intermittent deviations of the instantaneous throughput from the average throughput. These fluctuations coupled with the bursty nature of streaming video results in a situation where a burst is not serviced within the inter-burst time, causing a buffer underrun event. In order to minimize the fluctuation in instantaneous throughput, we increase MF in multiples of two until we observe stable performance.
    \begin{figure}[!t]
\includegraphics[width=0.5\textwidth]{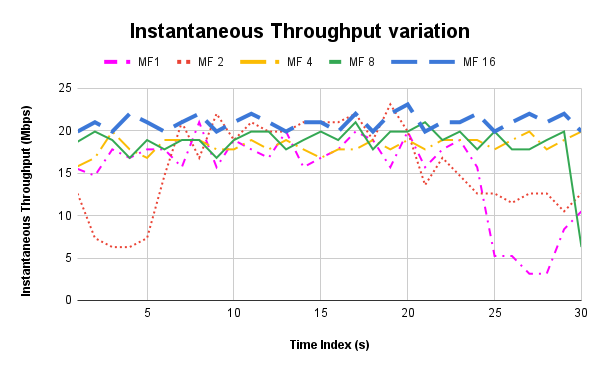} 
\caption{Instantaneous throughput variation for Different MF's.}
\label{fig:Instantaneous Graph}
\centering
\end{figure}
    
    \subsubsection{Phase 3} In this phase, we test the performance of the synthetic traffic model and calculate the corresponding QoS metrics. From Phase $1$ and $2$, we obtain the minimum required TWT duty cycle and MF required to maintain the QoS, and test the performance of the CBR synthetic traffic model. If the TWT schedule does not meet the QoS requirements, we increment the duty cycle by $5\%$ and reevaluate the performance. Once we obtain the TWT schedule for the CBR model, we test the performance of the VBR model.

\subsection{Results}
We generate video traffic (both CBR and VBR) at $15.6$ Mbps bit-rate, as mentioned in  Section \ref{for result}, and analyze the performance in terms of the required QoS metrics, i.e., average throughput and buffer health status. Throughout this section, \textbf{QoS-1} refers to the average throughput and \textbf{QoS-2} refers to the number of buffer underrun events (i.e., the number of times the buffer health status metric is not met) in each session. The set-up used for all the experiments is according to Fig. \ref{fig:testbed} where the radio parameters are mentioned in Table \ref{table:Set-up-parameters}. The performance of the system and the impact of adding the DUT in the network are also analyzed.
\begin{table}[!t]
\centering
  \begin{tabular}{|l|l|p{0.8 cm}|p{0.8 cm}|p{0.8 cm}|p{0.8 cm}|p{0.8 cm}|}
    \hline
    \multirow{3}{0.6 cm}{Traffic model} &
    \multirow{3}{1.2cm}{Performance without TWT}&
      \multicolumn{5}{c|}{Performance with TWT}\\ \cline{3-7}
      & {}  &Iteration 1 & Iteration 2 & Iteration 3 &Iteration 4 &  Iteration 5   \\
      \hline
   CBR&
61.2 &
66.3&
56.6&
62.4&
59.4&
61.4
 \\ 
 \hline
VBR &
56.8 &
62.8 &
61.0 &
65.6 &
64.2 &
58.1
 \\
 \hline

    \hline
  \end{tabular}
  \caption{System performance with and without TWT.}
\label{table:System Performance}
\end{table}

\subsubsection{Impact of TWT on the system performance}
In this section we observe the impact of adding the DUT (which is receiving a 15.6 Mbps CBR stream) to the network, in the presence of three iPerf sessions from the AP to Client 1, Client 2 and Client 3. Each iPerf session contains 8 parallel TCP streams (referred to as peak background congestion). The following table shows the performance of the system and the impact of a $30$\% TWT duty cycle on the net throughput of the system. Table \ref{table:System Performance} shows the sum of the throughput (in Mbps) of all iPerf sessions, not including the throughput of the synthetic traffic stream.



From Table \ref{table:System Performance}, we observe that even with a $30$\% TWT duty cycle, the net performance of the system does not deteriorate. Essentially, the AP uses the extra airtime to satisfy the demands of non-TWT clients.

\subsubsection{Impact of MF on synthetic traffic performance}
We evaluate the impact of varying MF on the performance of the CBR stream while keeping the TWT duty cycle constant. We run the experiment with peak background congestion. We define time spent in buffer underrun as time elapsed after the inter-burst time to serve out the burst. In Fig. \ref{fig:Buffer-underrun} we plot the total time spent in buffer underrun state over the duration of a simulation for different MF’s. From Fig. \ref{fig:Buffer-underrun}, we observe an increasing trend in performance, with the best performance for MF 8 followed by a sharp decrease thereafter. From Wireshark packet capture analysis, we note that when the value of MF is low, the corresponding TWT schedule has a large sleep time (TWT-WI) which delays TCP window size progression and leads to long round trip times at the tail end of the service periods. On the other hand, for higher MF values, the corresponding TWT schedule has a small wake duration (TWT-SP) which essentially hinders MAC layer frame aggregation, causing poor performance.   

\begin{figure}[!ht]
\includegraphics[width=0.5\textwidth]{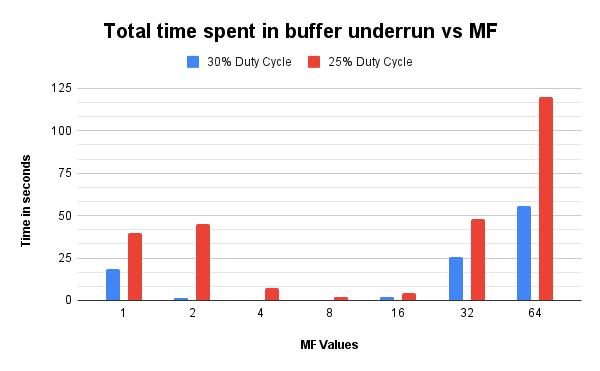}
\caption{Total time spent in buffer underrun state over the duration of a simulation for different MF’s.}
\label{fig:Buffer-underrun}
\centering
\end{figure}

\begin{table}[!t]
\centering
  \begin{tabular}{|l|l|l|l|l|}
    \hline
    \multirow{2.5}{*}{Iteration No.} &
      \multicolumn{4}{c|}{Duty cycle parameters}\\
      \cline{2-5}&
      \multicolumn{2}{c|}{25\% MF 8} &
      \multicolumn{2}{c|}{30\% MF 8} \\
      \cline{2-5}
      
      & {QoS 1} & {QoS 2} & {QoS 1} & {QoS 2}  \\
      \hline
     1  & 15 Mbps &5 &16 Mbps & 3\\ \hline
     2 & 15 Mbps & 5 & 16 Mbps & 2 \\ \hline

     3 & 16 Mbps & 2 &16 Mbps& 1 \\ \hline
     4 & 16 Mbps & 4 & 16 Mbps  & 0 \\ \hline
      5 & 15 Mbps & 4 & 16 Mbps & 1 \\ \hline
  \end{tabular}
  \caption{Performance comparison between $25$\% and $30$\% TWT duty cycle with MF $8$ on CBR stream}
\label{table:Duty_Cyle_Comparison}
\end{table}

\begin{table}[!t]
\centering
  \begin{tabular}{|l|l|l|l|l|}
    \hline
    \multirow{2.5}{*}{Iteration No.} &
      \multicolumn{4}{c|}{Traffic Model}\\ \cline{2-5}&
      \multicolumn{2}{c|}{CBR} &
      \multicolumn{2}{c|}{VBR} \\
      \cline{2-5}
      & {QoS 1} & {QoS 2} & {QoS 1} & {QoS 2}  \\
      \hline
     1  & 16 Mbps &3 &14 Mbps&2\\
    \hline
    2 &16 Mbps &2 &15 Mbps&0 \\
    \hline

    3 & 16 Mbps&1&14 Mbps&2 \\
    \hline
    4 & 16 Mbps& 0 & 15 Mbps & 3\\
     \hline
    5 & 16 Mbps&1
&
14 Mbps
&1

\\

    \hline
  \end{tabular}
  \caption{Performance comparison of VBR and CBR traffic model for $30$\% TWT duty cycle with MF $8$}
\label{table: CBR VBR Comparison}
\end{table}
\subsubsection{Performance variation for different TWT duty cycles}
From the Section \ref{phase1} we obtain a $30$\% TWT duty cycle with MF $8$ as the ideal choice to start testing the performance of the CBR stream. From Table \ref{table:Duty_Cyle_Comparison} we observe that it satisfies both the QoS requirements. These tests were performed with peak background congestion. Table \ref{table:Duty_Cyle_Comparison} summarizes the performance of CBR traffic for $25$\% and $30$\% TWT duty cycle with MF $8$. Each session of synthetic traffic lasted for about $2$ minutes.

\subsubsection{Performance Comparison of CBR and VBR traffic model}

  As described in Section \ref{VBR} the requirements of VBR traffic are more stringent than CBR traffic. In this experiment, both traffic sources emulate a $15.6$ Mbps bit-rate video with each streaming session running for about $2$ minutes. Table \ref{table: CBR VBR Comparison} summarizes the performance of both CBR and VBR models for $30$\% TWT duty cycle with MF $8$. There is no discernible  difference in the performance of both the synthetic traffic models with the TWT parameters obtained from our approach.

\section{Conclusion and Future Works}
In this work, we consider a network which comprises of a Wi-Fi $6$  AP, a TWT-enabled client, and multiple background non-TWT clients running at peak congestion.
\begin{figure}[!ht]
\includegraphics[width=0.4\textwidth]{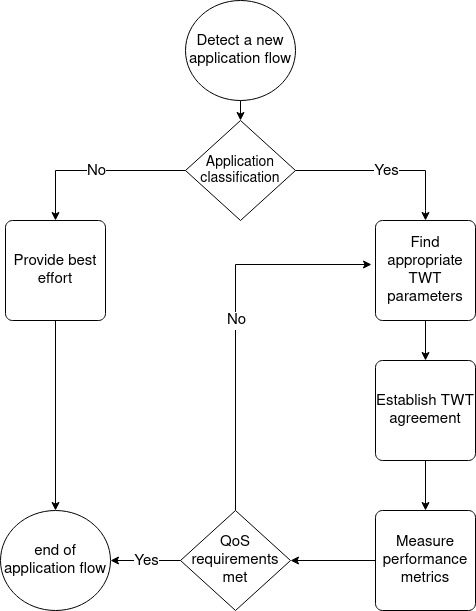}
\caption{Planned future work.}
\label{fig:Conclusion Flow Diagram}
\centering
\end{figure}
We analyze the performance evaluation metrics of synthetic traffic models that emulate streaming and show the effect of TWT on the network performance. Initial testing shows that key QoS metrics can be met for streaming traffic upon enabling TWT even in the presence of peak background congestion in the network. The unused airtime (i.e. the time remaining after servicing the TWT-enabled client) obtained from enabling TWT is utilized by the AP in servicing other non-TWT clients while satisfying the necessary QoS requirements of the TWT-enabled client. Since the objective of this work is to primarily schedule video streaming traffic in a congested network scenario, we do not measure the power consumption. 
Consequently by providing a TWT schedule, we indeed save power when compared to staying awake all the time while satisfying the QoS metrics. As an extension of this work, we plan to analyze different classes of applications and their respective QoS requirements. We also plan to evaluate the performance of an application and provide a suitable TWT schedule in order to satisfy the QoS needs of the application.
\newpage

\bibliographystyle{IEEEtran}
\bibliography{reference}

\begin{thebibliography}{10}
\providecommand{\url}[1]{#1}
\csname url@samestyle\endcsname
\providecommand{\newblock}{\relax}
\providecommand{\bibinfo}[2]{#2}
\providecommand{\BIBentrySTDinterwordspacing}{\spaceskip=0pt\relax}
\providecommand{\BIBentryALTinterwordstretchfactor}{4}
\providecommand{\BIBentryALTinterwordspacing}{\spaceskip=\fontdimen2\font plus
\BIBentryALTinterwordstretchfactor\fontdimen3\font minus
  \fontdimen4\font\relax}
\providecommand{\BIBforeignlanguage}[2]{{%
\expandafter\ifx\csname l@#1\endcsname\relax
\typeout{** WARNING: IEEEtran.bst: No hyphenation pattern has been}%
\typeout{** loaded for the language `#1'. Using the pattern for}%
\typeout{** the default language instead.}%
\else
\language=\csname l@#1\endcsname
\fi
#2}}
\providecommand{\BIBdecl}{\relax}
\BIBdecl

\bibitem{Standard}
``{{IEEE} Standard for Information Technology--Telecommunications and
  Information Exchange between Systems Local and Metropolitan Area
  Networks--Specific Requirements Part 11: Wireless LAN Medium Access Control
  (MAC) and Physical Layer (PHY) Specifications Amendment 1: Enhancements for
  High-Efficiency WLAN},'' \emph{IEEE Std 802.11ax-2021 (Amendment to IEEE Std
  802.11-2020)}, pp. 1--767, 2021.

\bibitem{Latency1}
\BIBentryALTinterwordspacing
E.~de~Carvalho~Rodrigues, A.~Garc{\'{\i}}a{-}Rodr{\'{\i}}guez, L.~G. Giordano,
  and G.~Geraci, ``On the latency of {IEEE} 802.11ax {WLAN}s with parameterized
  spatial reuse,'' 2020. [Online]. Available:
  \url{https://arxiv.org/abs/2008.07482}
\BIBentrySTDinterwordspacing

\bibitem{Bellata-TWT-Scheduled-Access}
M.~Nurchis and B.~Bellalta, ``Target wake time: Scheduled access in {IEEE}
  802.11ax {WLAN}s,'' \emph{IEEE Wirel. Commun.}, vol.~26, no.~2, pp. 142--150,
  2019.

\bibitem{DBLP:Bellata-journals/corr/abs-1811-00957}
\BIBentryALTinterwordspacing
K.~Dovelos and B.~Bellalta, ``Optimal resource allocation for uplink {OFDMA} in
  802.11ax networks,'' \emph{CoRR}, vol. abs/1811.00957, 2018. [Online].
  Available: \url{http://arxiv.org/abs/1811.00957}
\BIBentrySTDinterwordspacing

\bibitem{DBLP:Uplink-Resource-Allocation}
S.~Bhattarai, G.~Naik, and J.-M.~J. Park, ``Uplink resource allocation in ieee
  802.11ax,'' in \emph{ICC 2019 - 2019 IEEE International Conference on
  Communications (ICC)}, 2019, pp. 1--6.

\bibitem{ONLINE:Arista-White-Paper-MU-MIMO}
\BIBentryALTinterwordspacing
“Multi-User MIMO in Wi-Fi 6 - Arista Networks",. [Online]. Available:
  \url{https://www.arista.com/assets/data/pdf/Whitepapers/MU-MIMO-Whitepaper.pdf}
\BIBentrySTDinterwordspacing

\bibitem{ONLINE:WLAN-TGax-Evaluation-methodology}
\BIBentryALTinterwordspacing
“11ax Evaluation Methodology", IEEE 802.11-14/0571r12. [Online]. Available:
  \url{https://mentor.ieee.org/802.11/dcn/14/11-14-0571-12-00ax-evaluation-methodology.docx}
\BIBentrySTDinterwordspacing

\bibitem{mondal2017candid}
A.~Mondal, S.~Sengupta, B.~R. Reddy, M.~Koundinya, C.~Govindarajan, P.~De,
  N.~Ganguly, and S.~Chakraborty, ``Candid with youtube: Adaptive streaming
  behavior and implications on data consumption,'' in \emph{Proceedings of the
  27th Workshop on Network and Operating Systems Support for Digital Audio and
  Video}, 2017, pp. 19--24.

\bibitem{streaming_data}
\BIBentryALTinterwordspacing
``Streaming video - mbs to gbs - mobility report - ericsson.'' [Online].
  Available:
  \url{https://www.ericsson.com/en/reports-and-papers/mobility-report/articles/streaming-video}
\BIBentrySTDinterwordspacing

\bibitem{dash-cloudflair}
\BIBentryALTinterwordspacing
``What is {MPEG-DASH}? {HLS vs. DASH} | cloudflare.com.'' [Online]. Available:
  \url{https://www.cloudflare.com/learning/video/what-is-mpeg-dash/}
\BIBentrySTDinterwordspacing

\bibitem{zhang2017modeling}
T.~Zhang, F.~Ren, W.~Cheng, X.~Luo, R.~Shu, and X.~Liu, ``Modeling and
  analyzing the influence of chunk size variation on bitrate adaptation in
  {DASH},'' in \emph{IEEE INFOCOM 2017-IEEE Conference on Computer
  Communications}.\hskip 1em plus 0.5em minus 0.4em\relax IEEE, 2017, pp. 1--9.

\bibitem{7057917}
J.-M. Jeong and J.-D. Kim, ``Effective bandwidth measurement for dynamic
  adaptive streaming over http,'' in \emph{2015 International Conference on
  Information Networking (ICOIN)}, 2015, pp. 375--378.

\bibitem{WLAN-TGax-Evaluation-methodology}
\BIBentryALTinterwordspacing
“11ax Evaluation Methodology", IEEE 802.11-14/0571r12. [Online]. Available:
  \url{https://mentor.ieee.org/802.11/dcn/14/11-14-0571-12-00ax-evaluation-methodology.docx}
\BIBentrySTDinterwordspacing

\bibitem{ghobadi2012trickle}
M.~Ghobadi, Y.~Cheng, A.~Jain, and M.~Mathis, ``Trickle: Rate limiting {YouTube
  Video Streaming},'' in \emph{2012 USENIX Annual Technical Conference (USENIX
  ATC 12)}, 2012, pp. 191--196.

\end{thebibliography}

\end{document}